# Photochemical-induced phase transitions in photoactive semicrystalline polymers

Ruobing Bai, Eric Ocegueda, Kaushik Bhattacharya*

Division of Engineering and Applied Science, California Institute of Technology, Pasadena, California 91125, USA

* Corresponding author: bhatta@caltech.edu

**Abstract**

The emergent photoactive materials through photochemistry make it possible to directly convert photon energy to mechanical work. There is much recent work in developing appropriate materials and a promising new system is semi-crystalline polymers of the photoactive molecule azobenzene. We develop a phase field model with two order parameters for the crystal-melt transition and the *trans-cis* photo-isomerization to understand such materials, and the model describes the rich phenomenology. We find that the photo-reaction rate depends sensitively on temperature: at temperatures below the crystal-melt transition temperature, photoreaction is collective, requires a critical light intensity and shows an abrupt first order phase transition manifesting nucleation and growth; at temperatures above the transition temperature, photoreaction is independent and follows first order kinetics. Further, the phase transition depends significantly on the exact forms of spontaneous strain during the crystal-melt and *trans-cis* transitions. A non-monotonic change of photo-persistent *cis* ratio with increasing temperature is observed accompanied by a reentrant crystallization of *trans* below the melting temperature. A pseudo phase diagram is subsequently presented with varying temperature and light intensity along with the resulting actuation strain. These insights can assist the further development of these materials.

# I. INTRODUCTION

Light is a powerful energy source due to wireless transmission, intrinsic rich tunability, and potential high energy density using lasers. Directly converting photon energy to mechanical work is attractive, and has become the topic of interest in recent years thanks to emergent photoactive materials through either the photothermal or the photochemical effect (see a few recent reviews [1-4]). This paper focuses on the photochemical effect, where the photomechanical energy transduction is achieved through photoreaction (e.g., photo-isomerization and -dimerization) in materials embedded with photochromes. When illuminated with light of certain wavelength, the photochromes absorb light, transform to new configuration and change shape. A solid embedded with or composed of many such molecules deforms macroscopically under light and generates mechanical work if subject to an external load.

This photomechanical process involves an intrinsic coupling between the mechanics of the solid material and the photochemistry under illumination. Recent studies have revealed that in contrast to photoreactions in solution that undergo a first-order reaction kinetics, photoreactions in solids show a collective behavior through intermolecular interaction [5-8]. This collective behavior leads to phase transitions in the material, the details of which depend significantly on factors including the exact molecular interaction, elasticity, illumination, external load, temperature, and the energy landscape at various length scales.

Solid-state photomechanical actuation has been mainly achieved in liquid crystal polymers [1-3] and molecular crystals [4,9-11]. Photochromes can be embedded in a polymer network containing liquid crystal mesogens. Upon illumination, the transformation of photochromes changes the liquid crystalline order and thus the configuration of the polymer network, leading to large deformation. Photochromes can also be densely packed to form a crystal. The photoreaction directly changes the crystal structure and induces deformation. Both material systems have their advantages and drawbacks. For example, the relatively low photochrome concentration and the mainly entropic change of the polymer configuration in liquid crystal polymers generally limit their work output. On the other hand, despite the potentially large mechanical work output, photoactive molecular crystals mostly suffer poor processibility as well as fracture under illumination due to internal stresses [11,12].

A new class of materials, photoactive semi-crystalline polymers, have recently been developed [13] which exploit the light-induced crystallization and melting of photoactive crystals and polymers due to photoreaction [14-18]. The semi-crystalline to melt transition temperature of polymers of photoactive molecules like poly-azobenzene depends on the light-induced *trans* to *cis* ratio. Consequently, photo-isomerization can lead to light-induced melting which in turn can lead to large changes of shape. Such materials have the potential to harness the advantages of both liquid crystal polymers and molecular crystals, to generate large work output while maintaining a relatively simple synthesis with mechanical robustness. The goal of the current work is to develop a phase field model that can provide insight into the rich phenomenology of such materials due to the interplay between temperature, light, chemistry, and mechanics.

We start with the molecular structure of the photoactive semi-crystalline polymer in Section II.A. We then build a phase field model with two order parameters for the crystal-melt transition and the *trans-cis* photo-isomerization in the rest of Section II. We analyze the photo-persistent crystal ratio and *cis* ratio with different temperature and light intensity in Section III. A reentrant crystallization in the material is observed below the melting temperature. The photo-persistent states also depend on the heterogeneity, elasticity, and initial nuclei amount as shown in Section IV. We further present a pseudo phase diagram with varying temperature and light intensity in Section V, and the resulting actuation strains in Section VI. Finally, we provide additional discussions and conclude in Section VII.

## II. MOLECULAR STRUCTURE AND FORMULATION OF THE MODEL

### A. Molecular structure of photoactive semi-crystalline polymers

Following Kuenstler et al. [13], the molecular structure of the photoactive semi-crystalline polymer is illustrated in Fig. 1a. Azobenzene molecules are incorporated in the backbone of a polymer network to form a main-chain liquid crystal polymer. These molecules prefer the *trans* state (represented by a rod in Fig. 1a) in the absence of light but isomerize to the *cis* state (represented by a bent rod in Fig. 1a) under illumination of certain wavelength. Experiment shows that there are three phases: *trans*-(semi)crystal, *cis* melt and *trans* melt [13]. Illumination of the *trans*-semicrystalline phase and the consequent photo-isomerization can lead to a light-induced melting from the *trans*-semicrystalline phase to the *cis*-melt phase. The *trans*-melt phase can be obtained either by heating the *trans*-semicrystalline phase or thermal relaxation of the *cis*-melt phase.

**B. Landau free energy with two order parameters**

We develop a phase field model to study the phase transitions in the semi-crystalline polymer. Two order parameters are used as

$$\phi_1 = \begin{cases} 1 & \text{crystal,} \\ 0 & \text{melt.} \end{cases}$$
$$\phi_2 = \begin{cases} 1 & \text{cis,} \\ 0 & \text{trans.} \end{cases} \quad (1)$$

At temperature $T$, the bulk Landau free energy density of the system is expressed as

$$f_1(\phi_1,\phi_2,T) = A_1\phi_1^2(\phi_1-1)^2 + A_2\phi_2^2(\phi_2-1)^2 + A_3\phi_1^2\phi_2^2 + B(T-T_m)\phi_1^2 + C\left(\frac{1}{2}\phi_2^2 - \frac{1}{3}\phi_2^3\right), \quad (2)$$

where $A_1$, $A_2$, $A_3$, $B$, and $C$ are positive constants and $T_m$ is the crystal melting temperature. The first two terms describe the double well structures of the free energy corresponding to $\phi_1$ and $\phi_2$. The third term serves as an energetic penalization for the unfavorable *cis*-crystal phase. The fourth term governs the temperature-dependent first-order phase transition of $\phi_1$, and the fifth term introduces an energy difference between *trans* and *cis* following the single-molecule energy landscape of azobenzene [19]. The representative landscape of $f_1(\phi_1,\phi_2,T)$ at $T = T_m$ is shown in Fig. 1b. At this temperature the system has three minima corresponding to *trans*-crystal, *trans*-melt, and *cis*-melt, with *cis*-melt at a higher local minimum.

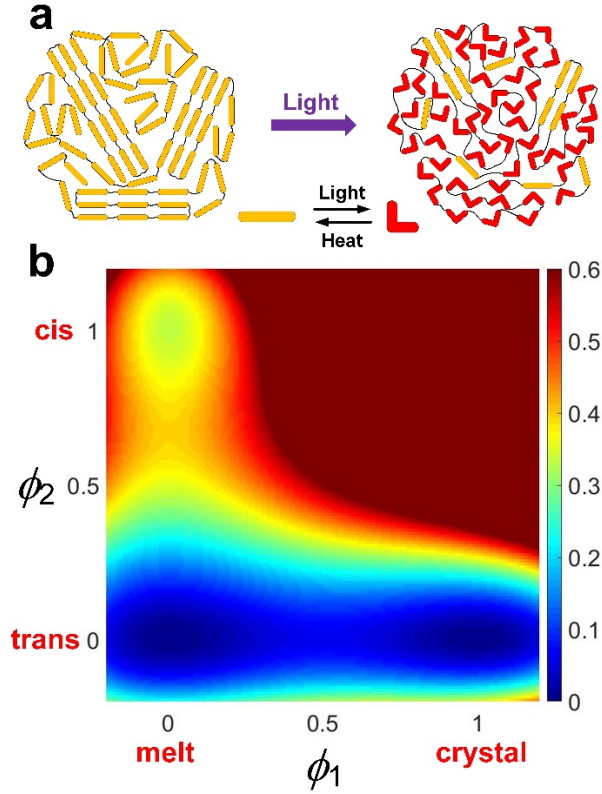

**FIG. 1.** Photoactive azobenzene semi-crystalline polymers. **(a)** Molecular structures of the semi-crystalline polymer incorporating azobenzene under illumination. The yellow (straight) and red (bent) rods represent azobenzene molecules in the *trans* and *cis* states, respectively. **(b)** A representative landscape of the bulk Landau free energy $f_1(\phi_1, \phi_2, T)$ at $T = T_m$.

## C. Elastic energy induced by phase transition

The phase transitions between the three phases is also accompanied by a change of shape when the stresses are zero. Therefore we introduce a state-dependent *spontaneous* or *stress-free* strain $\varepsilon_{ij}^*(\phi_1(\mathbf{x},t), \phi_2(\mathbf{x},t))$ expressed as

$$\varepsilon_{ij}^*(\phi_1, \phi_2) = (1-\phi_1)\phi_2 \varepsilon_{ij}^{*1} + \phi_1 \varepsilon_{ij}^{*2}, \tag{3}$$

where we have taken the *trans*-melt as the reference state, $\varepsilon_{ij}^{*1}$ is the spontaneous strain corresponding to *trans*-melt to *cis*-melt, and $\varepsilon_{ij}^{*2}$ corresponding to *trans*-melt to *trans*-crystal. We further assume $\varepsilon_{ij}^{*1} = diag(\varepsilon^{*1}, \varepsilon^{*1})$ and $\varepsilon_{ij}^{*2} = diag(\varepsilon^{*2}, -\varepsilon^{*2})$, indicating that the *trans-cis* isomerization induces a volumetric strain while the melt-crystal transition induces a deviatoric strain.

We assume (for simplicity) that the material has the same elastic constants before and after phase transitions, described by its stiffness tensor $C_{ijkl}$. The density of elastic energy is [20,21]

$$f_2(\varepsilon_{ij}, \phi_1, \phi_2) = \frac{1}{2} C_{ijkl} (\varepsilon_{ij} - \varepsilon_{ij}^*(\phi_1, \phi_2))(\varepsilon_{kl} - \varepsilon_{kl}^*(\phi_1, \phi_2)), \tag{4}$$

where $\varepsilon_{ij}(\mathbf{x}, t)$ is the strain field. For simplicity, the stiffness tensor is assumed to be isotropic homogeneous linear elastic:

$$C_{ijkl} = \lambda \delta_{ij} \delta_{kl} + \mu (\delta_{ik} \delta_{jl} + \delta_{il} \delta_{jk}), \tag{5}$$

where $\lambda = 2\mu\nu/(1-2\nu)$, $\nu$ is the Poisson's ratio, and $\mu$ is the shear modulus.

## D. Heterogeneity

There are heterogeneities such as non-photoactive functional groups in a semi-crystalline polymer, and these may affect the phase transitions. Therefore, we introduce an additional term [22,23]

$$H(\phi_1, \phi_2, \mathbf{x}) = H_0 [r_1(\mathbf{x})\phi_1 + r_2(\mathbf{x})\phi_2] \tag{6}$$

in the free energy where $H_0$ is the heterogeneity strength, and $r_1(\mathbf{x})$ and $r_2(\mathbf{x})$ are two random fields (quenched noise) with zero spatial average $\langle r_1(\mathbf{x}) \rangle = \langle r_2(\mathbf{x}) \rangle = 0$ and a prescribed correlation length $h$. We numerically generate these using a filtering method with fast Fourier transform (see Appendix A for the detailed algorithm). A representative pattern of the random fields is shown in Fig. 2a with a correlation length $h = 0.1$ along arbitrary directions.

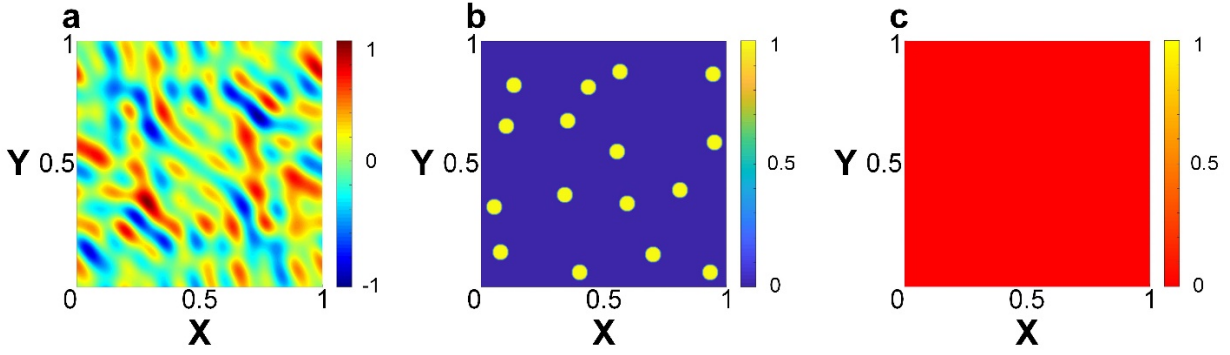

**FIG. 2. (a)** A representative pattern of random field $r(\mathbf{x})$ with zero spatial average and correlation length $h = 0.1$ along arbitrary directions. **(b)&(c)** The initial conditions ($t = 0$) of $\phi_1$ and $\phi_2$ in the square periodic domain, respectively.

### E. Total free energy

With all the above included, the bulk free energy density of the system is $f = f_1 + f_2 + H$. The total free energy of a volume $V$ is

$$F(\varepsilon_{ij}, \phi_1, \phi_2, T) = \int_V \left[ f(\varepsilon_{ij}, \phi_1, \phi_2, \mathbf{x}, T) + \frac{\lambda_1^2}{2}|\nabla \phi_1|^2 + \frac{\lambda_2^2}{2}|\nabla \phi_2|^2 \right] dV, \qquad (7)$$

in which we have included the exchange (Ginzburg) energies rising from gradients of the order parameters, which correspond to the interfacial energies between different phases [24].

### F. Kinetics of photoreaction

We study the kinetics of various phase transitions at constant temperature *T*. We consider two processes. The first is thermal relaxation with rates of $-\alpha_1(\delta F / \delta \phi_1)$ and $-\alpha_2(\delta F / \delta \phi_2)$, where $\alpha_1$ and $\alpha_2$ are positive and $\delta F$ is the first variation of *F*. The second is *trans-cis* photo-isomerization that follows a first order reaction kinetics with a rate of $\gamma I (1-\phi_1)(1-\phi_2)$, where $\gamma$ is positive constant, *I* the light intensity, and $(1-\phi_1)(1-\phi_2)$ the *trans* ratio in the melt. Therefore, $\phi_1$ and $\phi_2$ evolve according to the driven Allen-Cahn equation [25] or driven time-dependent Landau-Ginzburg equations [23]

$$\begin{cases} \dfrac{\partial \phi_1}{\partial t} = -\alpha_1 \dfrac{\delta F}{\delta \phi_1} = -\alpha_1 \left[ \dfrac{\partial f}{\partial \phi_1} - \lambda_1^2 \nabla^2 \phi_1 \right], \\ \dfrac{\partial \phi_2}{\partial t} = -\alpha_2 \dfrac{\delta F}{\delta \phi_2} + \gamma I (1-\phi_1)(1-\phi_2) = -\alpha_2 \left[ \dfrac{\partial f}{\partial \phi_2} - \lambda_2^2 \nabla^2 \phi_2 \right] + \gamma I (1-\phi_1)(1-\phi_2). \end{cases} \qquad (8)$$

The relaxation mobilities $\alpha_1$ and $\alpha_2$ are expected to increase with $T$. In particular, an increasing $T$ promotes the *cis*-to-*trans* backward reaction. While the dependence of these mobilities on $T$ are often described by either the Arrhenius-type behavior or the Williams–Landel–Ferry equation depending on the range of temperature applied [26,27], here we assume a linear relation since we are interested in a narrow temperature range:

$$\begin{cases} \alpha_1 (T/T_m) = \alpha_{1,0} \left( 1 + \beta_1 \dfrac{T}{T_m} \right), \\ \alpha_2 (T/T_m) = \alpha_{2,0} \left( 1 + \beta_2 \dfrac{T}{T_m} \right). \end{cases} \qquad (9)$$

where $\beta_1$ and $\beta_2$ are both positive constants.

### G. Balance of force

We assume that the evolution of the order parameters is slow compared to the propagation of sound through the system, and therefore the system is always in mechanical equilibrium. We can therefore solve for the strain $\varepsilon_{ij}$ at each time by solving the equation [20,21]

$$C_{ijkl} \dfrac{\partial \varepsilon_{kl}}{\partial x_j} = C_{ijkl} \dfrac{\partial \varepsilon_{kl}^* (\phi_1(\mathbf{x},t), \phi_2(\mathbf{x},t))}{\partial x_j} \qquad (10)$$

subject to appropriate boundary conditions. In this work we assume that the material is macroscopically stress-free such that

$$\langle \sigma_{ij} \rangle = \langle C_{ijkl} (\varepsilon_{kl} - \varepsilon_{kl}^* (\phi_1, \phi_2)) \rangle = 0, \qquad (11)$$

where $\langle \ \rangle$ denotes the spatial average in a periodic domain.

### H. Dimensionless groups and numerical implementation

We model the material system as a 2D periodic square domain with 512×512 nodes extended infinitely. The length of the square is unit so it normalizes all the relevant length scales in this problem. We define two time scales for the relaxation of $\phi_1$ and $\phi_2$: $\tau_1 = (\alpha_{1,0} A_1)^{-1}$ and $\tau_2 = (\alpha_{2,0} A_2)^{-1}$. We define two length scales for the feature sizes of the phase boundaries: $L_1 = \sqrt{\lambda_1^2 / A_1}$ and $L_2 = \sqrt{\lambda_2^2 / A_2}$. The dimensionless light intensity is $\Gamma = \gamma I \tau_2$. With these, the three governing equations of kinetics and force balance are simplified as

$$\begin{cases} \tau_1 \dfrac{\partial \phi_1}{\partial t} = -\left[ \dfrac{1}{A_1} \dfrac{\partial f(\varepsilon_{ij}, \phi_1, \phi_2, T)}{\partial \phi_1} - L_1^2 \nabla^2 \phi_1 \right] \left(1 + \beta_1 \dfrac{T}{T_m}\right), \\ \tau_2 \dfrac{\partial \phi_2}{\partial t} = -\left[ \dfrac{1}{A_2} \dfrac{\partial f(\varepsilon_{ij}, \phi_1, \phi_2, T)}{\partial \phi_2} - L_2^2 \nabla^2 \phi_2 \right] \left(1 + \beta_2 \dfrac{T}{T_m}\right) + \Gamma (1 - \phi_1)(1 - \phi_2), \\ C_{ijkl} \dfrac{\partial \varepsilon_{kl}}{\partial x_j} = C_{ijkl} \dfrac{\partial \varepsilon_{kl}^*(\phi_1, \phi_2)}{\partial x_j}. \end{cases} \quad (12)$$

We assign $\tau_1 = 1$ as the normalizing time, and $A_1 = 1$ as the normalizing energy density, such that both the bulk free energy density $f$ and stiffness tensor $C_{ijkl}$ are normalized by $A_1$. To simplify the further derivation including those in the Appendix, we redefine the parameters such that the dimensionless free energy is expressed as

$$f(\varepsilon_{ij}, \phi_1, \phi_2, \mathbf{x}, T) = \phi_1^2 (\phi_1 - 1)^2 + A_2 \phi_2^2 (\phi_2 - 1)^2 + A_3 \phi_1^2 \phi_2^2 + B_{T_m}\left(\dfrac{T}{T_m} - 1\right)\phi_1^2 + C\left(\dfrac{1}{2}\phi_2^2 - \dfrac{1}{3}\phi_2^3\right)$$
$$+ \dfrac{1}{2} C_{ijkl} \left(\varepsilon_{ij} - \varepsilon_{ij}^*(\phi_1, \phi_2)\right)\left(\varepsilon_{kl} - \varepsilon_{kl}^*(\phi_1, \phi_2)\right) + H_0 \left[ r_1(\mathbf{x}) \phi_1 + r_2(\mathbf{x}) \phi_2 \right]. \quad (13)$$

All the parameters are listed in Table 1 with their values given unless otherwise noted. The choices of values are based on a qualitative match with experimental observations reported in [13], as well as a focus on the scenario where different physical processes (e.g., light-induced reaction and temperature-induced relaxation) have comparable effects on the photochemical-induced phase transitions.

We comment that the characteristic length scale $L_2$ is smaller than our computational resolution. In general, this can lead to mesh dependence, pinning, and other artifacts. However, this is not so in our case.

We are concerned with a photo-persistent (almost steady) state while the difficulties with under-resolution typically distort the kinetics and not the steady state. In particular, the order parameter $\phi_2$ is driven by light and this driving can break any pinning. This is especially true since we have two order parameters and the evolution and the patterns are controlled by the other order parameter $\phi_1$ that is fully resolved.

To elaborate, consider the limit when $L_2$ goes to zero. Then, we would separate scales and minimize over all possible phase arrangements of $\phi_2$ at each pixel; i.e., use a mean-field theory where we replace $f$ with $f^{**}$ that is obtained by convexifying $f$ with respect to $\phi_2$. However, the energy $f$ is convex with respect to $\phi_2$ in the crystal phase (there is only the *trans* state), and only non-convex in the melt phase. Further, it is driven (by light) and thus driven out of the mixed *cis-trans* states; i.e., the solution remains in regions where $f^{**} = f$. This is exactly what happens when we formally set $L_2 = 0$ in our equations.

**Table 1. Parameters in the phase field model**

| Dimensionless parameter | Value | Physical meaning |
|---|---|---|
| $A_2$ | 3 | Parameters in the Landau free energy normalized by $A_1$ ($A_1 = 1$) |
| $A_3$ | 3 | |
| $B$ | 1 | |
| $C$ | 2 | |
| $\mu$ | 0 (default), 1, 10 | Shear modulus |
| $\nu$ | 0.25 | Poisson's ratio |
| $H_0$ | 0, 0.2 (default), 0.4 | Strength of the heterogeneity noise |
| $h$ | 0.1 | Correlation length of the heterogeneity noise |
| $\varepsilon^{*1}$ | -0.1 | Magnitude of volumetric spontaneous strain |
| $\varepsilon^{*2}$ | 0.1 | Magnitude of deviatoric spontaneous strain |
| $L_1$ | 0.01 | Size of phase boundary $L_1 = \sqrt{\lambda_1^2 / A_1}$ |
| $L_2$ | $10^{-5}$ | Size of phase boundary $L_2 = \sqrt{\lambda_2^2 / A_2}$ |
| $\tau_1$ | 1 (normalizing time) | Relaxation time of $\phi_1$ |
| $\tau_2$ | 0.1 | Relaxation time of $\phi_2$ |
| $\delta t$ | 0.01 | Time step of the finite difference calculation |

| $\Gamma$ | 1.6 (default) or varying | Normalized light intensity |
|---|---|---|
| $\beta_1$ | 0.1 | Temperature-mobility coefficient for $\phi_1$ |
| $\beta_2$ | 1 | Temperature-mobility coefficient for $\phi_2$ |
| $T/T_m$ | Varying | Normalized temperature |

The details of the numerical implementation are described in Appendix B. In short, we adopt a finite difference algorithm with a hybrid implicit-explicit method. We express terms involving the gradient and Laplace operators implicitly, and the rest terms explicitly. We then solve the partial differential equations using the fast Fourier transform (FFT) and a fixed Lagrangian (material) grid (the linear elasticity we adopt in the current model further does not distinguish between the Lagrangian and Eulerian grids). In solving the equation of force balance involving the field of photo-strain $\varepsilon_{ij}^*$, we follow the theoretical result derived in our previous work [7]. The time step of the finite difference calculation is set to be $\delta t = 0.01$.

Unless otherwise noted, the initial conditions are shown in Fig. 2b&c, where the system contains 4×4 randomly distributed circular nuclei of crystal phase ($\phi_1 = 1$) with diameter of 1/16 surrounded by the melt phase ($\phi_1 = 0$), and a homogeneous field of *trans* state ($\phi_2 = 0$).

We solve our evolution equations for a long time ($t_{max} = 50$ in non-dimensional units). The microstructure evolves quickly initially but then reaches a state where it is almost stationary (e.g. from Movies 1 and 2 [28]). We call this state the *photo-persistent state*. While the microstructure reaches an equilibrium in many cases, it is not so in all cases, especially when it involves mixed states. In such a situation, the state persists for a very long time, but can suddenly change. This is consistent with theoretical understanding, other simulations and experimental observations [29,30]. Since these states persist for a very long time, we consider them to be representative of experimental observation. In any case, while specific details may depend on $t_{max}$, the qualitative features are independent of the particular choice of $t_{max}$.

We could have directly studied the steady state equation associated with Eq. (8) where the left-hand-side is zero. However, this may have multiple solutions and it is not immediately clear which solutions are accessible by a kinetic process. We also remark that the equilibrium state is not necessarily the thermodynamic equilibrium. The system may be caught in a local minimum. More importantly, we have a driven system under continuous light illumination that couples to only one of the two order parameters. That is, the driving term $\gamma I(1-\phi_1)(1-\phi_2)$ is present in only one of the two equations in Eq. (8). Thus, we cannot recast Eq. (8) as a gradient flow of an effective free energy $F_\gamma$: i.e. write the system in the form $\partial \phi_1 / \partial t = -\alpha_1 \delta F_\gamma / \delta \phi_1$, $\partial \phi_2 / \partial t = -\alpha_2 \delta F_\gamma / \delta \phi_2$, and then study the minima of $F_\gamma$.

The photo-persistent state in many of our simulations involve inhomogeneous microstructures. We do not always show the detailed microstructure of every simulation result, but their existence in the photo-persistent state can be identified through the average order parameters $<\phi_1>$ and $<\phi_2>$. When $<\phi_1>$ and $<\phi_2>$ are close to either 0 or 1, the system is rather homogeneous with minimum microstructures. When $<\phi_1>$ and $<\phi_2>$ are between 0 and 1, inhomogeneous microstructures exist in the system.

## III. TEMPERATURE-LIGHT DEPENDENT PHOTO-PERSISTENT STATES

We first explore how photo-persistent state depends on light intensity at two different temperatures in Fig. 3. Below the melting temperature, the photo-persistent state is *trans*-crystal in the absence of any illumination as shown in Fig. 3a. Surprisingly, this photo-persistent state remains unchanged – in particular there is no *trans* to *cis* isomerization and the material remains in the fully *trans* state – for low light intensity below a critical intensity. Beyond this critical intensity, there is significant light-induced melting as the *cis* fraction increases rapidly and crystal fraction decreases rapidly with increasing illumination. The reason for the suppression of the *trans*-to-*cis* isomerization below a critical light intensity is that azobenzene molecules cannot isomerize independently due to the intermolecular interactions in the crystalline state. The molecules therefore behave collectively, and there is not enough light to trigger a collective isomerization. This collective behavior at low temperature due to the larger intermolecular interaction has been observed experimentally in azobenzene self-assembly monolayers [31] and molecular crystals [6], and has been recently shown in our study on photoactive solids using an Ising spin model [7].

The situation is different at high temperatures above the melting temperature as shown in Fig. 3b. The material is a *trans*-melt at zero illumination. As the light intensity increases, so does the fraction of *cis* molecules, initially slowly but then rapidly. The material is always in the melt.

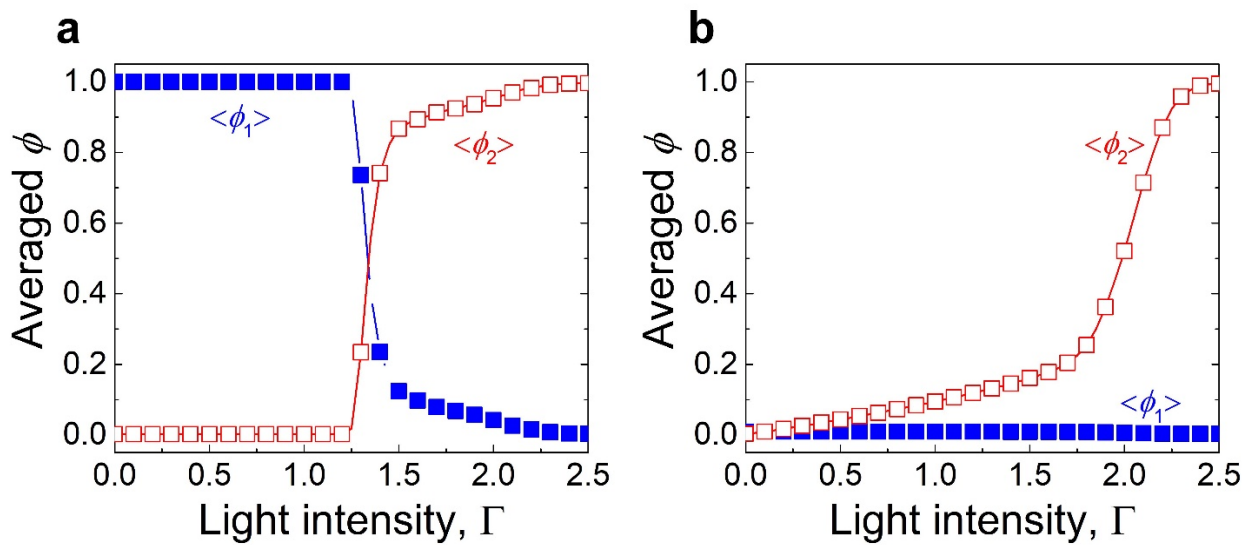

**FIG. 3.** Photo-persistent states vs light intensity at low temperature and high temperature, $H_0 = 0.2$. $<\phi_1>$: solid blue squares. $<\phi_2>$: open red squares. **(a)** At low temperature $T/T_m = 0.6$, the discontinuity of $<\phi_1>$ and $<\phi_2>$ indicates a first-order phase transition. **(b)** At high temperature $T/T_m = 1.2$, the rather smooth curve of $<\phi_2>$ indicates a first-order reaction kinetics.

We next fix the light intensity to be $\Gamma = 1.6$ and study the dependence of photo-persistent $<\phi_1>$ and $<\phi_2>$ on increasing temperature. As shown in Fig. 4a, the curves show a non-monotonic dependence on temperature due to a competition between the temperature-induced melting, the light-induced *trans*-to-*cis* isomerization, and the temperature-induced *cis*-to-*trans* backward relaxation. At low temperature (e.g., $T/T_m = 0.4$), the system is semi-crystalline and the melt phase nearly completely isomerizes to *cis* under the constant illumination $\Gamma = 1.6$, but the light does not further melt the *trans*-crystal. With increasing temperature, the same illumination melts more crystalline domains and transforms them to *cis*-melt. As a result, the average crystal ratio $<\phi_1>$ decreases and the average *cis* ratio $<\phi_2>$ increases. Further increasing the temperature also increases the mobility of the thermal relaxation, to a point where the *cis*-to-*trans* backward relaxation starts to dominate, leading to a decrease of $<\phi_2>$. If this takes place below the crystal melting temperature $T_m$, some slight reentrant crystallization of *trans* happens, as seen in Fig. 4a around $T/T_m = 0.9$. Fig. 4b shows the photo-persistent microstructures at different temperatures corresponding to those in Fig. 4a [28].

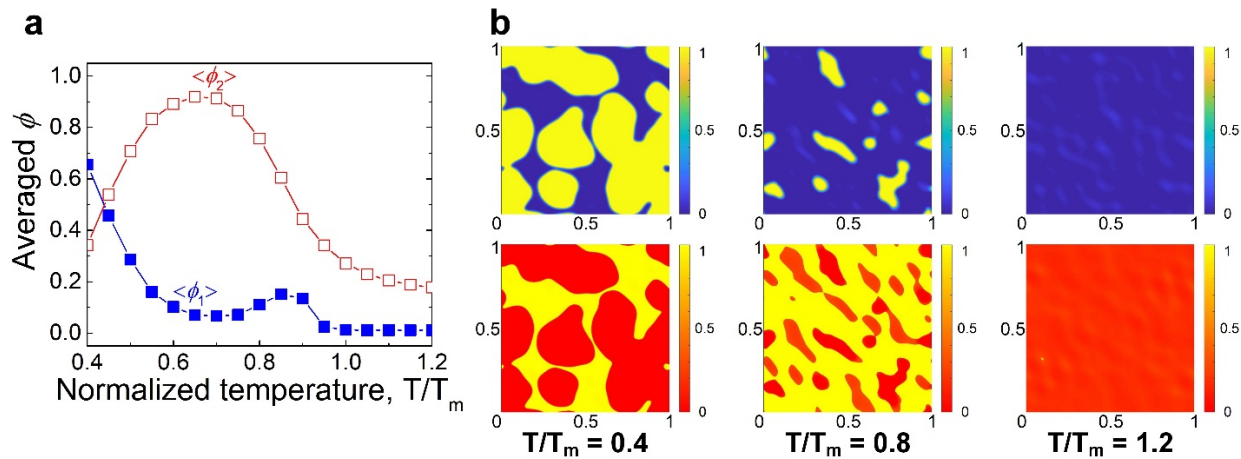

**FIG. 4. (a)** The photo-persistent $<\phi_1>$ and $<\phi_2>$ vs temperature with a constant illumination $\Gamma = 1.6$, $H_0 = 0.2$. The curves show non-monotonic dependence on the temperature. $<\phi_1>$: solid blue squares. $<\phi_2>$: open red squares. **(b)** The photo-persistent microstructures at different temperatures corresponding to those in Fig. 4a. Top row: $\phi_1$. Bottom row: $\phi_2$.

## IV. EFFECTS FROM HETEROGENEITY, ELASTICITY, AND INITIAL NUCLEI

There are non-photoactive heterogeneities in the semi-crystalline polymer such as the chain extender, crosslinks, residual solvents or initiators, and other precipitates. These heterogeneities can play important roles in the evolution of phase transition through pinning the phase boundaries with possible stick-slip behaviors [32,33]. We include the effect of this heterogeneity by a quenched spatial noise described in Section II.D, and now examine it in more detail. We set the spatial correlation length of the heterogeneity noise to be $h = 0.1$, much larger than the two interface length scales $L_1 = 0.01$ and $L_2 = 10^{-5}$. In this case, previous study has suggested that the pinning of phase boundary by the heterogeneity can be strong [32].

Fig. 5a shows the photo-persistent $<\phi_1>$ and $<\phi_2>$ vs temperature with different heterogeneity strength $H_0$. A larger $H_0$ smoothens the curves, shifting them from abrupt first-order like transition ($<\phi_1>$ and $<\phi_2>$ stay at 0 or 1) to gradual second-order like transition ($<\phi_1>$ and $<\phi_2>$ stay at intermediate values

3/1/2021 14

between 0 and 1). This is a direct consequence of the pinning of phase boundaries. As the heterogeneity strength increases, the pinning becomes stronger, leaving multiple domains of different phases coexisting in the system. The photo-persistent $<\phi_1>$ and $<\phi_2>$ therefore average out over many domains and show the intermediate values.

We then study the effect of elasticity on the phase transitions arising from the spontaneous strains. These may induce non-negligible stress and strain fields in mixed states, and the resulting elastic energy affects the phase transitions in return. We keep all the parameters unchanged from before except the shear modulus $\mu$. As shown in Fig. 5b, an increasing $\mu$ shifts the curves towards the abrupt first-order like transition ($<\phi_1>$ and $<\phi_2>$ stay at 0 or 1). Such a trend has also been found recently in our study on photoactive solids using an Ising spin model [7].

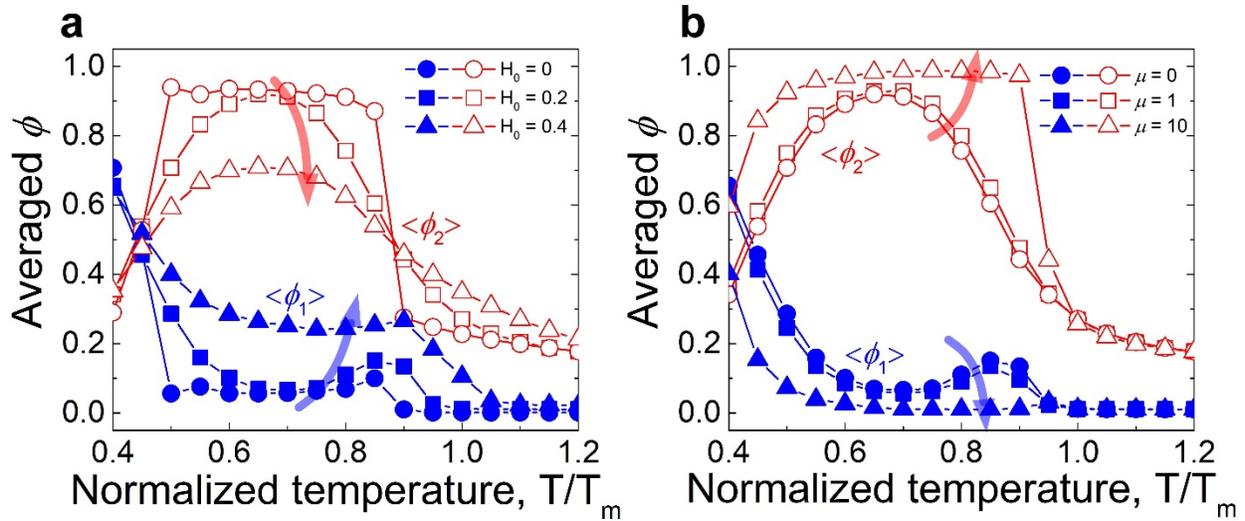

**FIG. 5. (a)** Photo-persistent $<\phi_1>$ and $<\phi_2>$ vs temperature with different heterogeneity strength $H_0$. The increase of $H_0$ is indicated by the direction of arrows. **(b)** Photo-persistent $<\phi_1>$ and $<\phi_2>$ vs temperature with different shear modulus $\mu$. The increase of $\mu$ is indicated by the direction of arrows. The constant illumination is $\Gamma = 1.6$. $<\phi_1>$: solid blue dots. $<\phi_2>$: open red dots.

Elasticity further affects through the exact forms of spontaneous strains. As shown in Eq. (3), we assume a volumetric spontaneous strain of magnitude $\varepsilon^{*1}$ for the *trans-cis* isomerization and a deviatoric spontaneous strain of magnitude $\varepsilon^{*2}$ for the melt-crystal transition. We now consider three cases with different values of ($\varepsilon^{*1}$, $\varepsilon^{*2}$) to be (-0.1, 0.1), (0, 0.1), and (-0.1, 0), and show their representative microstructures of $\phi_1$ in Fig. 6. The *trans*-crystal phase shows vertically aligned, diagonally aligned, and isotropic in the three cases respectively. The average crystal ratio $<\phi_1>$ at $t$ = 50 is slightly lower in the last case, possibly indicating a slower kinetics.

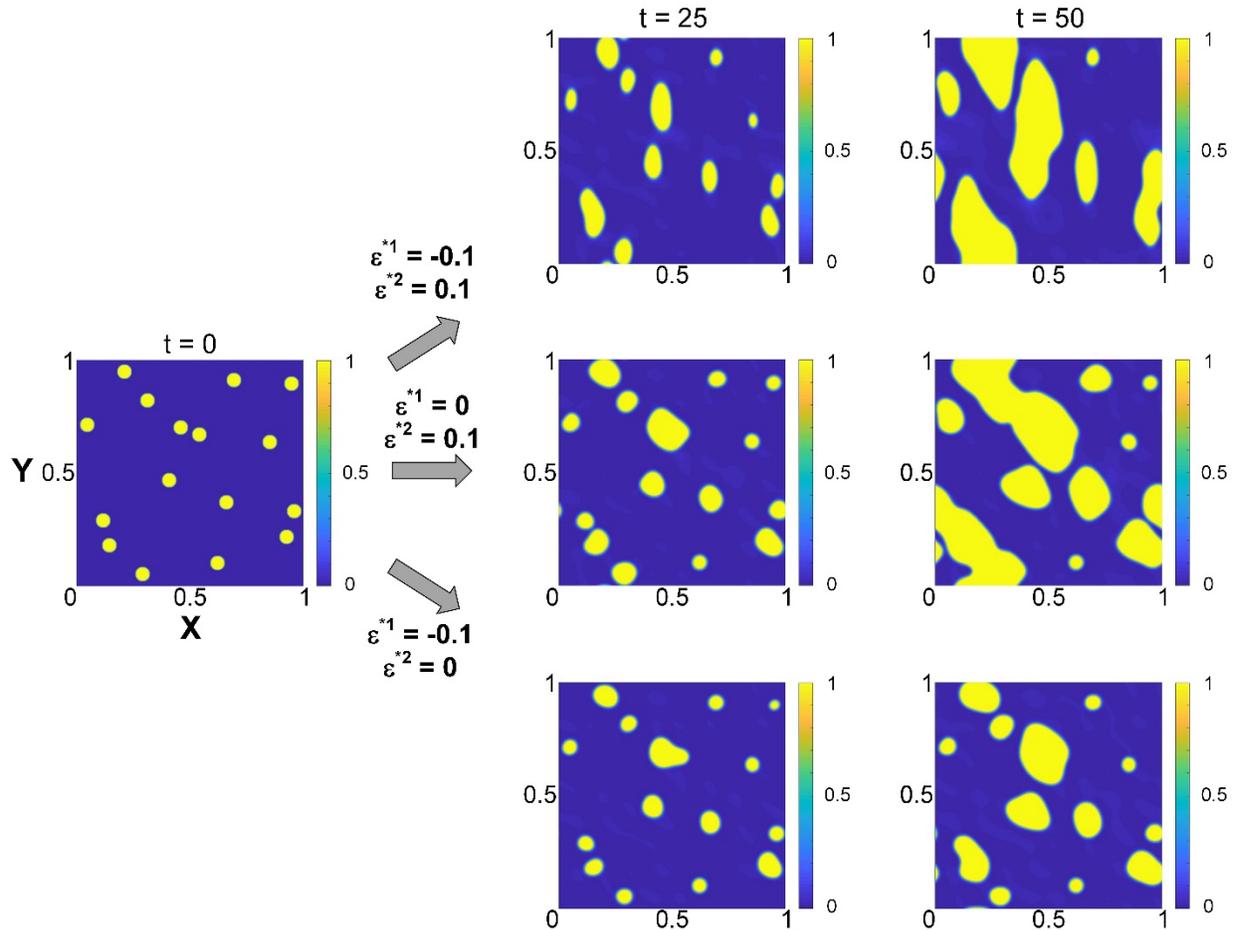

**FIG. 6.** Microstructures of *trans*-crystal ($\phi_1$ = 1, yellow islands) at $t$ = 0 (first column), $t$ = 25 (second column) and $t$ = 50 (third column) with different values of ($\varepsilon^{*1}$, $\varepsilon^{*2}$) to be (-0.1, 0.1), (0, 0.1), and (-0.1, 0). The *trans*-crystal phase shows vertically aligned, diagonally aligned, and isotropic in the three cases. The shear modulus is set to be $\mu$ = 10.

To explain the observed microstructures and kinetics with different ($\varepsilon^{*1}$, $\varepsilon^{*2}$), we recall Eq. (3) to calculate the total spontaneous strain tensor $\varepsilon^*_{ij}$. After examining the phase formations during the simulation (e.g. from Movies 1 and 2 [28]), one can find that the two main coexisting phases in each case of Fig. 6 are the *trans*-crystal phase ($\phi_1$ = 1, $\phi_2$ = 0) and the *cis*-melt phase ($\phi_1$ = 0, $\phi_2$ = 1). This is due to combined effects from the relatively strong illumination $\Gamma = 1.6$, the much smaller *trans-cis* relaxation time $\tau_2 = 0.1$ compared to the crystal-melt relaxation time $\tau_1 = 1$, and the smaller *trans-cis* interfacial energy indicated by $L_2 = 0.001$. The spontaneous strain tensors corresponding to these two phases are $diag(\varepsilon^{*2}, -\varepsilon^{*2})$ and $diag(\varepsilon^{*1}, \varepsilon^{*1})$, respectively. In the three cases shown in Fig. 6, the spontaneous strain tensors are calculated to be

$$\begin{cases} \text{Case 1} \left(\varepsilon^{*1} = -0.1, \varepsilon^{*2} = 0.1\right): diag(0.1, -0.1) \text{ and } diag(-0.1, -0.1). \\ \text{Case 2} \left(\varepsilon^{*1} = 0, \varepsilon^{*2} = 0.1\right): diag(0.1, -0.1) \text{ and } diag(0, 0). \\ \text{Case 3} \left(\varepsilon^{*1} = -0.1, \varepsilon^{*2} = 0\right): diag(0, 0) \text{ and } diag(-0.1, -0.1). \end{cases} \quad (14)$$

Recall the Hadamard jump condition [34,35], which states that two domains with strain tensors $\boldsymbol{\varepsilon}^+$ and $\boldsymbol{\varepsilon}^-$ can coexist across a boundary with normal $\mathbf{n}$ if there exists a vector $\mathbf{a}$ satisfying

$$\boldsymbol{\varepsilon}^+ - \boldsymbol{\varepsilon}^- = \frac{1}{2}(\mathbf{a} \otimes \mathbf{n} + \mathbf{n} \otimes \mathbf{a}). \quad (15)$$

Examining this condition for the three cases, we can find $\mathbf{n} = (1,0)^T$ for Case 1, $\mathbf{n} = (\pm 1, 1)^T$ for Case 2, but no existing $\mathbf{n}$ and $\mathbf{a}$ for Case 3. As a result, in the first two cases, neighboring domains can coexist with zero stress to ameliorate the total elastic energy if their phase boundary has the normal $\mathbf{n}$ found above, exactly as shown in Fig. 6. In the third case, because there is no such $\mathbf{n}$ that satisfies (15), internal stresses build up and the nucleation and growth is impeded by the increasing elastic energy. The phase domain growth proceeds relatively slower in an isotropic manner compared to the other cases, which explains the slower kinetics.

It is worthwhile noting that the current model assumes a constant elastic modulus before and after any phase transition. This is a simplifying assumption given the possibly large modulus difference between different phases. Nevertheless, the current analysis still provides a good qualitative understanding on the role of elastic interaction in photochemical-induced phase transitions.

The photoreaction and phase transition further depend on the initial amount of nuclei in the system as shown in Fig. 7. We increase the number of initial nuclei of *trans*-crystal from 4×4 to 8×8 (Fig. 7a), and compare the photo-persistent curves in Fig. 7b. An increase of <$\phi_1$> and decrease of <$\phi_2$> are observed with increasing number of initial crystal nuclei, due to the facilitated growth. In addition, a more evident reentrant crystallization at the intermediate temperature is observed with 8×8 initial nuclei.

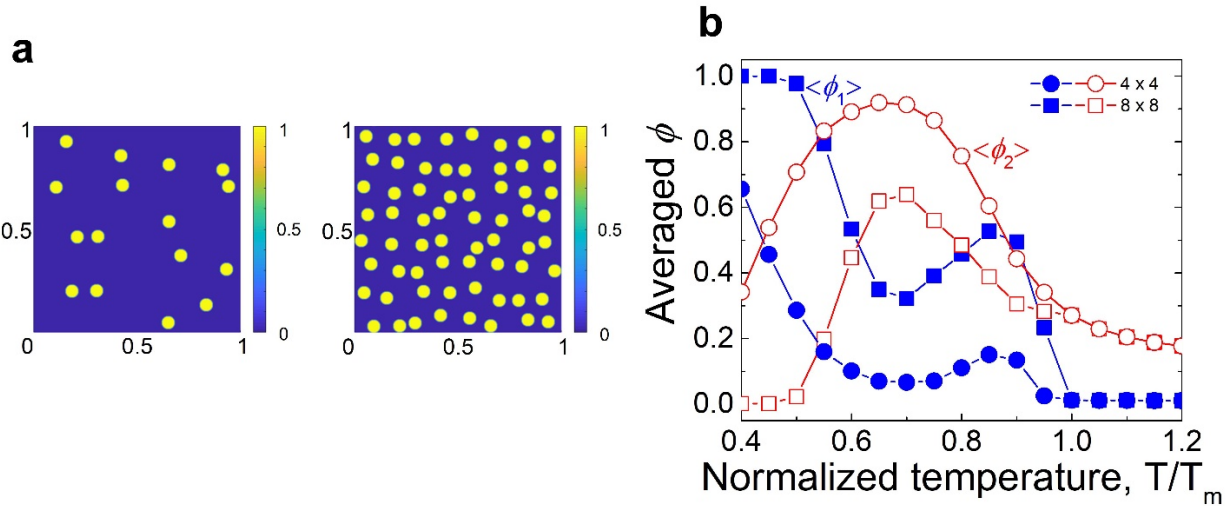

**FIG. 7.** The photoreaction and phase transition depend on the initial amount of nuclei in the system. **(a)** Two cases with the number of initial nuclei of *trans*-crystal (yellow islands) to be 4×4 (left) to 8×8 (right). **(b)** The photo-persistent <$\phi_1$> and <$\phi_2$> vs temperature comparing the two cases. <$\phi_1$>: solid blue dots. <$\phi_2$>: open red dots.

## V. REENTRANT CRYSTALLIZATION AND PSEUDO PHASE DIAGRAM

Next, we take a closer look at the reentrant crystallization below the melting temperature observed in Fig. 4a and Fig. 7b. We compute the photo-persistent <$\phi_1$> and <$\phi_2$> with different temperature and light intensity using an initial condition of 8×8 *trans*-crystal nuclei, and plot them in Fig. 8a and 8b. Phase boundaries are observed with kinked shapes, and are largely blurred as a consequence of mixed domains. The phase boundaries become sharper when we reduce the heterogeneity strength $H_0$ (results not shown here).

Combining Fig. 8a and 8b we construct a pseudo phase diagram in Fig. 8c by assigning phases based on the following rule: melt ($\langle\phi_1\rangle < 0.5$), crystal ($\langle\phi_1\rangle > 0.5$), *trans* ($\langle\phi_2\rangle < 0.5$), and *cis* ($\langle\phi_2\rangle > 0.5$). We emphasize that this *pseudo* phase diagram is different from common phase diagrams which are drawn from thermodynamic equilibrium because the illumination makes it a driven system. Notice the "nose" on the right of Fig. 8c which describes the reentrant crystallization: if we take an illumination around Γ = 1.6 and increase temperature, the *trans*-crystal first melts to the *cis*-melt, but then recrystallizes to the *trans*-crystal before transforming to the *trans*-melt. This reentrant crystallization with increasing temperature takes place within a narrow range of light intensity about Γ = 1.5-1.6. It is a result of a competition between the light-induced *trans*-to-*cis* isomerization and the thermal-induced *cis*-to-*trans* backward reaction: there is a region below the melting temperature $T_m$ where the thermal-induced backward reaction dominates over the light-induced isomerization.

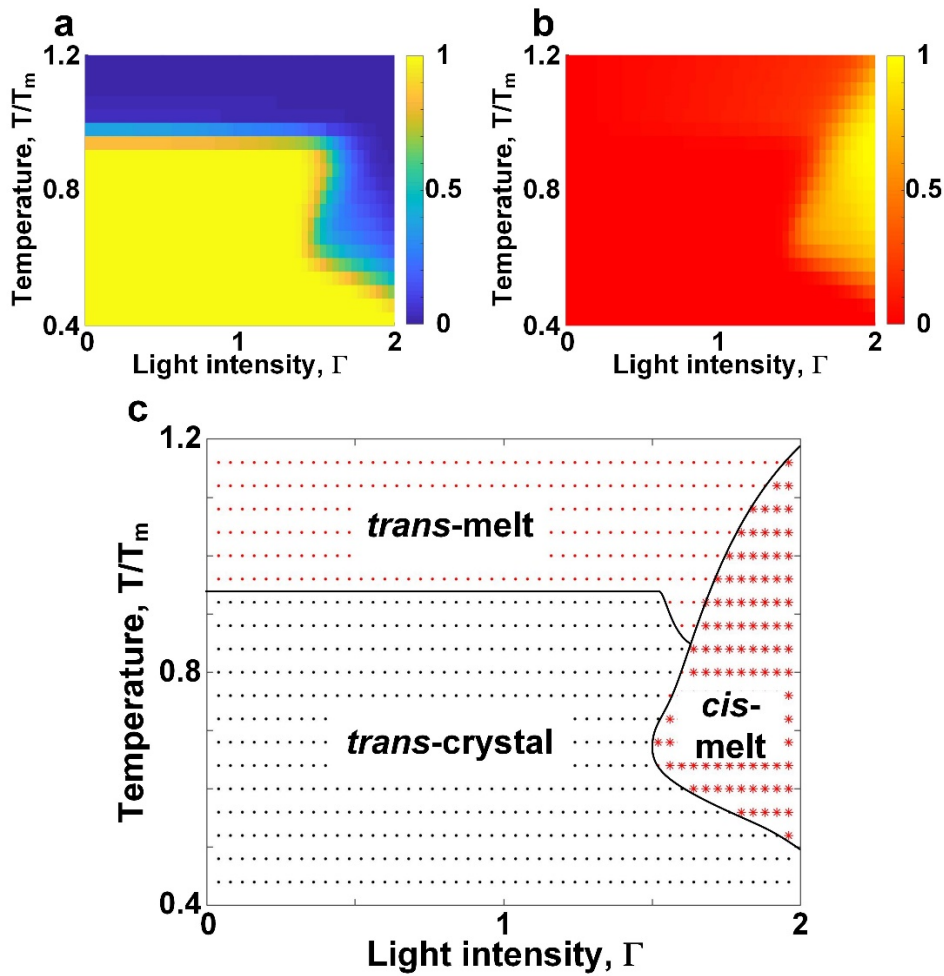

**FIG. 8.** Pseudo phase diagram of the semi-crystalline polymers, $\mu = 0$. **(a)&(b)** The photo-persistent <$\phi_1$> and <$\phi_2$> with different temperature and light intensity, respectively. **(c)** A pseudo phase diagram constructed by manually assigning phases based on the following rule: melt ($\langle\phi_1\rangle < 0.5$), crystal ($\langle\phi_1\rangle > 0.5$), *trans* ($\langle\phi_2\rangle < 0.5$), and *cis* ($\langle\phi_2\rangle > 0.5$).

This reentrant crystallization vanishes when we set the elastic modulus to be as large as $\mu = 10$. The "nose" disappears in the pseudo phase diagram shown in Fig. 9c. This observation is consistent with Fig. 5b, where an increasing $\mu$ (equivalently an increasing long-range elastic interaction) suppresses the transformation of individual molecules but promotes collective behavior. This possibly explains the reason that the reentrant crystallization has not yet been observed experimentally in photomechanical semi-crystalline polymers.

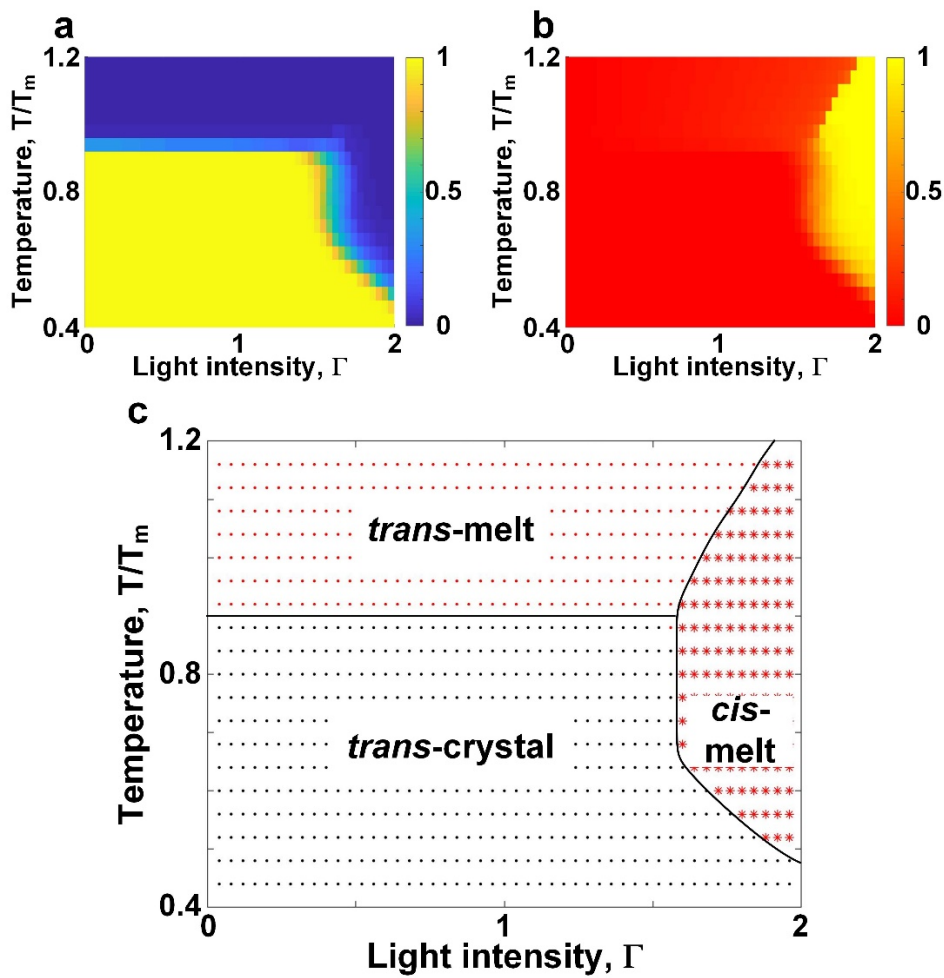

**FIG. 9.** Pseudo phase diagram of the semi-crystalline polymers, $\mu = 10$. **(a)&(b)** The photo-persistent $\langle\phi_1\rangle$ and $\langle\phi_2\rangle$ with different temperature and light intensity, respectively. **(c)** A pseudo phase diagram constructed by manually assigning phases based on the following rule: melt ($\langle\phi_1\rangle < 0.5$), crystal ($\langle\phi_1\rangle > 0.5$), *trans* ($\langle\phi_2\rangle < 0.5$), and *cis* ($\langle\phi_2\rangle > 0.5$).

## VI. ACTUATION STRAIN

To further investigate the photomechanical actuation performance, we plot the spatially averaged actuation strains at the photo-persistent state as a function of temperature and illumination in Fig. 10 with $\mu = 0$ and $\mu = 10$ (corresponding to the pseudo phase diagrams in Fig. 8 and 9). The average shear strain $\langle \varepsilon_{12} \rangle$ is nearly zero for the entire phase diagram, and so we plot the average hydrostatic (volumetric) strain $\langle \varepsilon_{11} + \varepsilon_{22} \rangle / 2$ and deviatoric (shear) strain $\langle \varepsilon_{11} - \varepsilon_{22} \rangle / 2$. All the actuation strains plotted here are calculated under a macroscopically stress-free far-field condition in (11).

The *trans*-melt phase is taken as the reference state, so both $\langle \varepsilon_{11} + \varepsilon_{22} \rangle / 2$ and $\langle \varepsilon_{11} - \varepsilon_{22} \rangle / 2$ are nearly zero in this phase in Fig. 10. The phase transition from *trans*-melt to *trans*-crystal induces nearly no hydrostatic strain but a deviatoric strain of approximately 0.1. By contrast, the phase transition from *trans*-melt to *cis*-melt induces a hydrostatic strain of approximately -0.1 but nearly no deviatoric strain. Finally, the phase transition from *trans*-crystal to *cis*-melt induces both a hydrostatic strain and a deviatoric strain of approximately -0.1. All these actuation strains are consistent with the form of spontaneous strain $\varepsilon_{ij}^*$ in (3). In addition, the reentrant crystallization can be observed in Fig. 10a with $\mu = 0$, where the actuation strain varies corresponding to the phase transition.

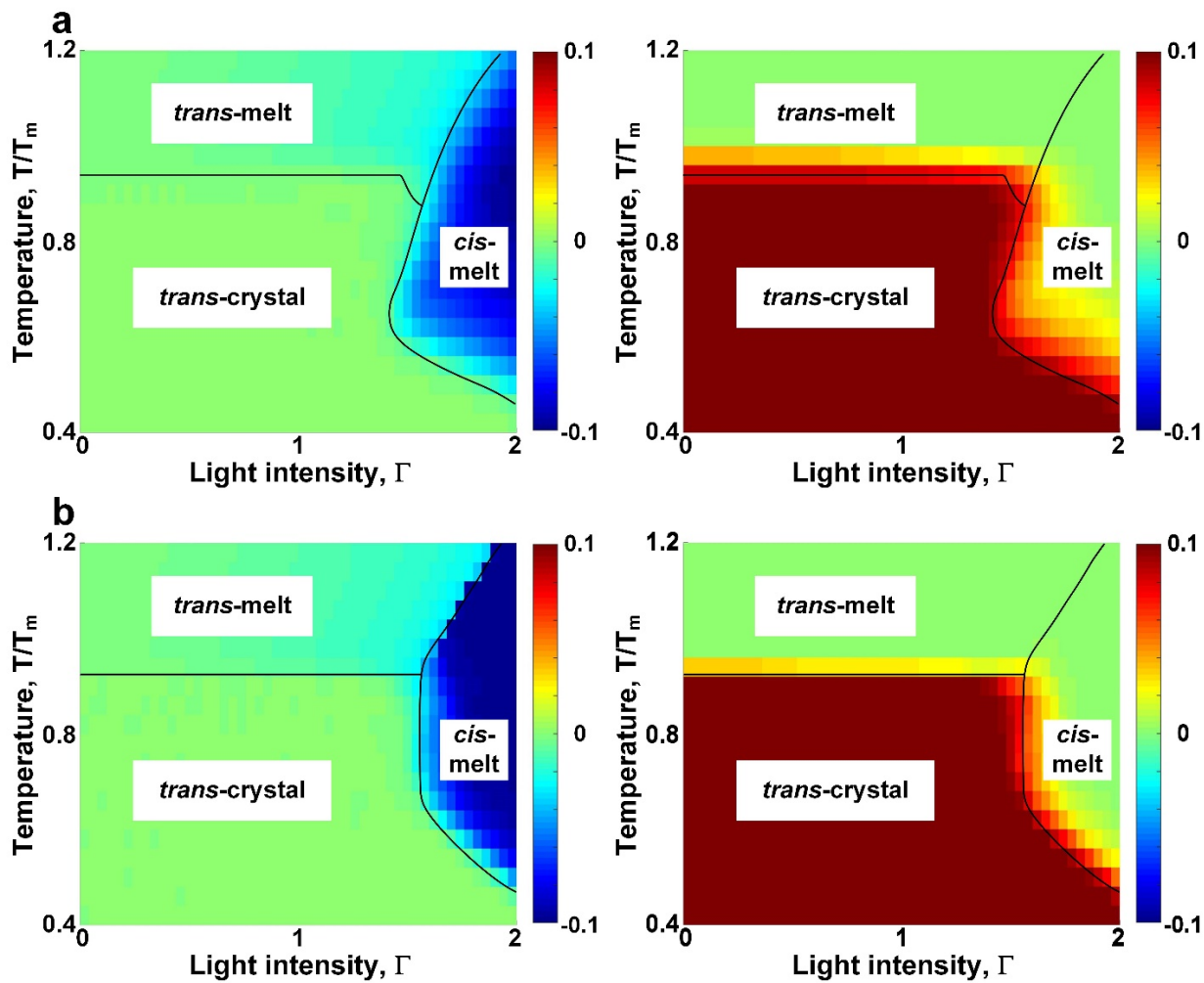

**FIG. 10.** Actuation strains of the semi-crystalline polymers with **(a)** $\mu = 0$ and **(b)** $\mu = 10$. Left: the photo-persistent average hydrostatic (volumetric) strain $\langle \varepsilon_{11} + \varepsilon_{22} \rangle / 2$. Right: the photo-persistent deviatoric (shear) strain $\langle \varepsilon_{11} - \varepsilon_{22} \rangle / 2$.

## VII. DISCUSSION AND CONCLUSION

Actuation through photochemical-induced phase transition has two advantages. First, unlike most actuation mechanisms that are limited to certain class of materials with a narrow range of mechanical property, photochemical phase transition takes place in a wide range of materials from densely packed molecular crystals to loosely crosslinked photoactive polymers. This high diversity not only enables numerous applications, but also raises interesting fundamental questions that vary among different material systems but have intrinsic correlations through photoreaction. Second and more importantly, phase transitions in materials especially first-order or steep second-order transitions can enable giant actuation (e.g., deformation, stroke, work) with an increment of small external stimulus. This idea has been actively explored in recent years in actuations using mechanical instability (e.g., snap-through [36], pull-in [37], buckling [38]), and also serves as the intrinsic mechanism of many thermal-induced actuations involving phase transition, such as shape memory alloys [39], shape memory polymers [40], temperature-sensitive hydrogels [41], and thermotropic liquid crystal polymers [42]. Compared to these materials, photochemical-induced phase transition does not rely on the relatively slow thermal conduction, which makes fast actuation promising. One such example is the recently developed high-frequency (~10-100 Hz) oscillating beams under illumination [43-45].

With these advantages above, we believe the mechanistic studies such as those in the current paper make it possible to further employ various phase transitions in photoactive materials for optimal performance. As an example, one can place a material at a state close enough to the phase boundary in the pseudo phase diagram, and subsequently stimulate it across the phase boundary to induce large change of shape or material property. This instability-like actuation can be integrated into pre-designed structures to push forward even more extreme actuations through structural instability. The degree of freedom involved in tuning the photochemical/photomechanical phase transitions is nearly infinite, considering the near-infinite degrees of freedom in mechanical stress states, light profiles, and material components.

Multiple scientific challenges still remain in both the modeling and experiment of existing photomechanical material systems. Available micro/meso-scale modeling frameworks for photomechanical behaviors are still limited, and quantitative comparison to experiment remains difficult. Models at the macroscopic or continuum level have shown some recent success in predicting experiments [45-47], but they are limited to special types of materials or geometries. The current work together with many recent studies are an important step to the development of a generalized modeling framework, by starting from microscopic theories such as statistical mechanics, towards meso- and macroscopic theories such as continuum mechanics, and eventually to structure mechanics at large scale. For experimental validation, most model systems such as azobenzene often suffer from small light penetration into the material [45] as well as the wide overlapping absorption peaks in the spectrums of isomers [13,48], leading to low thermodynamic efficiency. These issues may be resolved with the development of new types of photochromes, such as the *negative photochrome* that allows more photons to penetrate upon photoreaction [49]. Other experimental challenges include quantitative identification and control of material parameters, reproducibility of material samples, and control of undesirable factors such as dissipation and fracture. Overall, the combination of modeling and experiment remains challenging even at the qualitative level, which call for an urgent investigation in the future.

The current model assumes linear elasticity with small deformation. While the strains considered here (±0.1) are slightly larger than what is usually considered as small deformation, the experience in martensitic and other phase transitions involving shape-change is that this is adequate for these strain range when one is dealing with overall behavior (e.g., [50,51]). The model may readily be extended to finite deformations by replacing Eq. (10) with the finite deformation analogy (e.g., [52]).

In the current model, we have chosen a Landau free energy with qualitative parameter values. These parameters can be further quantified through the combination of more detailed experiment and simulation. The double-well structure of the *trans-cis* energy landscape can be quantified through first-principle calculations of the single-molecule. The parameters related to the temperature dependent crystal-melt Landau free energy can be quantified through measuring material constants such as the melting temperature and the latent heat, using experiments such as differential scanning calorimetry. The remaining parameters relevant to the coupling between the two order parameters may be further obtained through more sophisticated experiments such as quantifying the photoreaction products under various temperature and light intensity. The selected values of the parameters in Table 1 were based on such simulations and experiments done previously [13,19].

Some of the qualitative predictions from the current model can be validated by experiments. For example, the light and temperature dependency of photo-persistent states can be quantitatively measured from the distinct absorption spectra of *trans* and *cis*, together with differential scanning calorimetry to measure the crystallinity. The observation of crystallization and the reentrant crystallization can be done through the polarizing microscopy. The detailed microstructure during the phase transition and in the photo-persistent state can be observed in the similar way. Some qualitative simulation results indeed show a relatively good match with experimental observations following the system reported in [13], which is currently in preparation and to be published later.

In summary, we have developed a phase field model with two order parameters to explore the photochemical-induced phase transition in photoactive semi-crystalline polymers. We examine the photo-persistent states under different temperature and light intensity. We find a first-order reaction kinetics at high temperature, but a first-order phase transition at low temperature. We also find a reentrant crystallization when illuminating and increasing the temperature from below the melting temperature. The photo-persistent states depend on the heterogeneity, elasticity, and the initial amount of nuclei. These couplings lead to a pseudo phase diagram of photochemical-induced phase transition along with the resulting actuation strains. We hope the study in this paper can help the development of future photomechanical materials by connecting the microscopic information to the macroscopic theory.

**ACKNOWLEDGEMENTS**

This work was motivated by the experimental work of Alexa S. Kuenstler, Hantao Zhou and Ryan Hayward, and we are grateful to them for numerous interesting discussions. We are also grateful for the support of the Office of Naval Research through the MURI on Photomechanical Material Systems (ONR N00014-18-1-2624).

**APPENDIX A: GENERATION OF THE HETEROGENEITY NOISE**

We generate the random heterogeneity noise field using a filtering method with fast Fourier transform (FFT). The noise field is targeted to have a dimensionless correlation length $h$ normalized by the length of the periodic square domain. To do so, we first generate a complex random field $(1+i)r(\mathbf{k})$ in the Fourier space, where $r(\mathbf{k})$ is a real random field $\in [-1,1]$. Subsequently, we use the Gaussian function to filter out all the Fourier components other than those at the vicinity of $k_1^2 + k_2^2 = k_0^2$, where $k_0 = 1/h$ is the reciprocal of the correlation length. In this way, we only keep the Fourier components that have correlation lengths of approximately $h$ in arbitrary directions in the real space. We then conduct the inverse FFT of the filtered random field and keep the real parts to obtain the correlated random field in the real space. Finally we rescale to generate the quenched noise with an amplitude $H_0$.

**APPENDIX B: ALGORITHM FOR THE PHASE FIELD MODELING**

The equations we solve are expressed in Eq. (12) with the free energy in Eq. (13). We use finite difference method to solve these three nonlinear PDEs with periodic boundary conditions. The time step in the finite difference is $\delta t$. For simplicity we write $\tau_{1t} = \tau_1 / \delta t$ and $\tau_{2t} = \tau_2 / \delta t$. we treat terms involving the gradient and Laplace operators implicitly, and the rest terms explicitly. The approximated PDEs become

$$\begin{cases} \tau_{1t}(1+\beta_1 T_{Tm})^{-1}(\phi_1^{n+1} - \phi_1^n) = -\left[\frac{\partial f_1}{\partial \phi_1} + \frac{\partial f_2}{\partial \phi_1} + H_0 r_1\right]\bigg|_n + L_1^2 \nabla^2 \phi_1^{n+1}, \\ \tau_{2t}(1+\beta_2 T_{Tm})^{-1}(\phi_2^{n+1} - \phi_2^n) = -\frac{1}{A_2}\left[\frac{\partial f_1}{\partial \phi_2} + \frac{\partial f_2}{\partial \phi_2} + H_0 r_2\right]\bigg|_n + L_2^2 \nabla^2 \phi_2^{n+1} + \Gamma(1+\beta_2 T_{Tm})^{-1}(1-\phi_1^n)(1-\phi_2^n), \\ C_{ijkl}\frac{\partial \varepsilon_{kl}^{n+1}}{\partial x_j} = C_{ijkl}\left[\frac{\partial \varepsilon_{kl}^*(\phi_1, \phi_2)}{\partial x_j}\right]\bigg|_n, \end{cases} \quad \text{(B1)}$$

where $n$ denotes the $n^{th}$ time step and $T_{Tm} = T/T_m$. We can further rewrite the first two equations by separating the implicit and explicit terms and get

$$\begin{cases} \nabla^2 \phi_1^{n+1}(\mathbf{x}) - G_1 \phi_1^{n+1}(\mathbf{x}) = G_2^n(\mathbf{x}) \\ \nabla^2 \phi_2^{n+1}(\mathbf{x}) - G_3 \phi_2^{n+1}(\mathbf{x}) = G_4^n(\mathbf{x}) \end{cases}, \quad \text{(B2)}$$

where

$$G_1 = \frac{\tau_{1t}(1+\beta_1 T_{Tm})^{-1}}{L_1^2}, \quad \text{(B3)}$$

$$G_2^n(\mathbf{x}) = \frac{1}{L_1^2}\left[\frac{\partial f_1}{\partial \phi_1} + \frac{\partial f_2}{\partial \phi_1} + H_0 r_1\right]\bigg|_n - \frac{1}{L_1^2}\tau_{1t}(1+\beta_1 T_{Tm})^{-1}\phi_1^n, \quad \text{(B4)}$$

$$G_3 = \frac{1}{L_2^2}\tau_{2t}\left(1+\beta_2 T_{Tm}\right)^{-1},\tag{B5}$$

$$G_4^n(\mathbf{x}) = \frac{1}{L_2^2}\frac{1}{A_2}\left[\frac{\partial f_1}{\partial \phi_2}+\frac{\partial f_2}{\partial \phi_2}+H_0 r_2\right]\bigg|_n - \frac{1}{L_2^2}\tau_{2t}\left(1+\beta_2 T_{Tm}\right)^{-1}\phi_2^n - \frac{1}{L_2^2}\Gamma\left(1+\beta_2 T_{Tm}\right)^{-1}\left(1-\phi_1^n\right)\left(1-\phi_2^n\right),\tag{B6}$$

$$\frac{\partial f_1}{\partial \phi_1} = 2(\phi_1)^2(2\phi_1-3)+2A_3(\phi_2)^2\phi_1+2\left[1+B(T_{Tm}-1)\right]\phi_1,\tag{B7}$$

$$\frac{\partial f_1}{\partial \phi_2} = 2A_2(\phi_2)^2(2\phi_2-3)-C(\phi_2)^2+2A_3(\phi_1)^2\phi_2+(C+2A_2)\phi_2,\tag{B8}$$

$$\frac{\partial f_2}{\partial \phi_1} = -C_{ijkl}\left(\varepsilon_{ij}-\varepsilon_{ij}^*(\phi_1,\phi_2)\right)\frac{\partial \varepsilon_{kl}^*}{\partial \phi_1},\tag{B9}$$

$$\frac{\partial f_2}{\partial \phi_2} = -C_{ijkl}\left(\varepsilon_{ij}-\varepsilon_{ij}^*(\phi_1,\phi_2)\right)\frac{\partial \varepsilon_{kl}^*}{\partial \phi_2}.\tag{B10}$$

**Solving PDEs in the Fourier space**

The general form of the first two PDEs after discretization is

$$\nabla^2\phi^{n+1}(\mathbf{x})-G_\alpha\phi^{n+1}(\mathbf{x})=G_\beta^n(\mathbf{x}),\tag{B11}$$

where $\phi$ stands for either $\phi_1$ or $\phi_2$, $G_\alpha$ is a constant scalar, and $G_\beta^n(\mathbf{x})$ is a scalar field which is determined by $\phi^n(\mathbf{x})$ in the previous step. Rewriting the PDE in Eq. (B11) in the Fourier space, we obtain

$$-k^2\tilde{\phi}^{n+1}(\mathbf{k})-G_\alpha\tilde{\phi}^{n+1}(\mathbf{k})=\tilde{G}_\beta^n(\mathbf{k}).\tag{B12}$$

This linear algebraic equation can be solved for

$$\tilde{\phi}^{n+1}(\mathbf{k})=\frac{\tilde{G}_\beta^n(\mathbf{k})}{-k^2-G_\alpha}.\tag{B13}$$

We obtain the function $\phi^{n+1}(\mathbf{x})$ in real space by inverse Fourier transform.

Following the same principle for the third PDE

$$C_{ijkl}\frac{\partial \varepsilon_{kl}^{n+1}}{\partial x_j}=C_{ijkl}\left[\frac{\partial \varepsilon_{kl}^*(\phi_1,\phi_2)}{\partial x_j}\right]\bigg|_n,\tag{B14}$$

we can first write the equation in the Fourier space

$$C_{ijkl}(ik_j)\tilde{\varepsilon}_{kl}^{n+1} = C_{ijkl}(ik_j)\tilde{\varepsilon}_{kl}^{*n}. \tag{B15}$$

We then use the equation $\tilde{\varepsilon}_{kl}^{n+1} = (ik_l\tilde{u}_k^{n+1} + ik_k\tilde{u}_l^{n+1})/2$ to represent the strain by displacement $u_i$, and solve for the displacement

$$\tilde{u}_i^{n+1}(\mathbf{k}) = \frac{-i}{\mu}\tilde{M}_{kl}^n\left[\frac{k_k}{k^2}\delta_{il} - \frac{1}{2(1-\nu)}\frac{k_ik_kk_l}{k^4}\right], \text{ for } \mathbf{k} \neq 0, \tag{B16}$$

where $\tilde{M}_{kl}^n = C_{klpq}\tilde{\varepsilon}_{pq}^{*n}$. The strain field is then

$$\tilde{\varepsilon}_{ij}^{n+1}(\mathbf{k}) = \frac{1}{2}(ik_j\tilde{u}_i^{n+1} + ik_i\tilde{u}_j^{n+1})$$
$$= \frac{1}{2\mu}\left(\tilde{M}_{ki}^n\frac{k_jk_k}{k^2} + \tilde{M}_{kj}^n\frac{k_ik_k}{k^2}\right) - \frac{1}{2\mu(1-\nu)}\frac{k_ik_jk_kk_l}{k^4}\tilde{M}_{kl}^n, \text{ for } \mathbf{k} \neq 0. \tag{B17}$$

For $\mathbf{k} = 0$, the PDE becomes trivial. Therefore, we need to solve this special case relying on the far-field condition. By definition, we know

$$\tilde{\varepsilon}_{ij}^{n+1}(\mathbf{k} = 0) = \int_V \varepsilon_{ij}^{n+1}(\mathbf{x})d\mathbf{x} = V\langle\varepsilon_{ij}^{n+1}\rangle, \tag{B18}$$

where $\langle\ \rangle$ denotes the spatial average inside the periodic volume $V$. For the current problem, we apply a zero-stress far-field condition, namely

$$\langle C_{ijkl}(\varepsilon_{kl} - \varepsilon_{kl}^*(\phi_1,\phi_2))\rangle = 0. \tag{B19}$$

This leads to

$$\langle\varepsilon_{kl}\rangle = \langle\varepsilon_{kl}^*\rangle. \tag{B20}$$

Writing it in the discretized form, we get

$$\tilde{\varepsilon}_{ij}^{n+1}(\mathbf{k} = 0) = V\langle\varepsilon_{ij}^{n+1}\rangle = V\langle\varepsilon_{ij}^{*n}\rangle. \tag{B21}$$

To sum up, the solved strain field is

$$\varepsilon_{ij}^{n+1}(\mathbf{x}) = \frac{1}{V}\sum_{\mathbf{k}\neq 0}\tilde{\varepsilon}_{ij}^{n+1}(\mathbf{k})\exp(i\mathbf{k}\cdot\mathbf{x}) + \frac{1}{V}\tilde{\varepsilon}_{ij}^{n+1}(\mathbf{k} = 0)$$
$$= \frac{1}{V}\sum_{\mathbf{k}\neq 0}\tilde{\varepsilon}_{ij}^{n+1}(\mathbf{k})\exp(i\mathbf{k}\cdot\mathbf{x}) + \langle\varepsilon_{ij}^{*n}\rangle, \tag{B22}$$

with

$$\tilde{\varepsilon}_{ij}^{n+1}(\mathbf{k}) = \frac{1}{2\mu}\left(\tilde{M}_{ki}^{n}\frac{k_j k_k}{k^2} + \tilde{M}_{kj}^{n}\frac{k_i k_k}{k^2}\right) - \frac{1}{2\mu(1-\nu)}\frac{k_i k_j k_k k_l}{k^4}\tilde{M}_{kl}^{n}, \text{ for } \mathbf{k} \neq 0. \tag{B23}$$

**A specific spontaneous strain field**

The algorithm up to now applies to arbitrary spontaneous strain field $\varepsilon_{ij}^{*}(\phi_1, \phi_2)$. We now calculate more detailed expressions based on the specific spontaneous strain field in the paper

$$\varepsilon_{ij}^{*}(\phi_1, \phi_2) = \psi^1(\mathbf{x})\varepsilon_{ij}^{*1} + \psi^2(\mathbf{x})\varepsilon_{ij}^{*2}, \tag{B24}$$

where $\psi^1 = (1-\phi_1)\phi_2$, $\psi^2 = \phi_1$, $\varepsilon_{ij}^{*1} = \begin{bmatrix} \varepsilon^{*1} & \\ & \varepsilon^{*1} \end{bmatrix}$ and $\varepsilon_{ij}^{*2} = \begin{bmatrix} \varepsilon^{*2} & \\ & -\varepsilon^{*2} \end{bmatrix}$.

The stiffness tensor of an isotropic homogeneous linear elastic material is

$$C_{ijkl} = \lambda \delta_{ij}\delta_{kl} + \mu(\delta_{ik}\delta_{jl} + \delta_{il}\delta_{jk}). \tag{B25}$$

Substituting all of these into $M_{ij} = C_{ijkl}\varepsilon_{kl}^{*}$ and $\tilde{M}_{ij} = C_{ijkl}\tilde{\varepsilon}_{kl}^{*}$, we obtain

$$M_{ij}(\mathbf{x}) = \begin{bmatrix} \frac{2\mu}{1-2\nu}\psi^1(\mathbf{x})\varepsilon^{*1} + 2\mu\psi^2(\mathbf{x})\varepsilon^{*2} & \\ & \frac{2\mu}{1-2\nu}\psi^1(\mathbf{x})\varepsilon^{*1} - 2\mu\psi^2(\mathbf{x})\varepsilon^{*2} \end{bmatrix}, \tag{B26}$$

and

$$\tilde{M}_{ij}(\mathbf{k}) = \begin{bmatrix} \frac{2\mu}{1-2\nu}\tilde{\psi}^1(\mathbf{k})\varepsilon^{*1} + 2\mu\tilde{\psi}^2(\mathbf{k})\varepsilon^{*2} & \\ & \frac{2\mu}{1-2\nu}\tilde{\psi}^1(\mathbf{k})\varepsilon^{*1} - 2\mu\tilde{\psi}^2(\mathbf{k})\varepsilon^{*2} \end{bmatrix}. \tag{B27}$$

Further substituting these into the express of strain in Eq. (B23), we obtain (for $\mathbf{k} \neq 0$)

$$\tilde{\varepsilon}_{11}^{n+1}(\mathbf{k}) = \frac{1}{\mu}\tilde{M}_{11}^{n}\frac{k_1^2}{|\mathbf{k}|^2} - \frac{1}{2\mu(1-\nu)}\left[\frac{k_1^4}{|\mathbf{k}|^4}\tilde{M}_{11}^{n} + \frac{k_1^2 k_2^2}{|\mathbf{k}|^4}\tilde{M}_{22}^{n}\right], \tag{B28}$$

$$\tilde{\varepsilon}_{22}^{n+1}(\mathbf{k}) = \frac{1}{\mu}\tilde{M}_{22}^{n}\frac{k_2^2}{|\mathbf{k}|^2} - \frac{1}{2\mu(1-\nu)}\left[\frac{k_2^4}{|\mathbf{k}|^4}\tilde{M}_{22}^{n} + \frac{k_1^2 k_2^2}{|\mathbf{k}|^4}\tilde{M}_{11}^{n}\right], \tag{B29}$$

$$\tilde{\varepsilon}_{12}^{n+1}(\mathbf{k}) = \frac{1}{2\mu} \frac{k_1 k_2}{|\mathbf{k}|^2} \left( \tilde{M}_{11}^n + \tilde{M}_{22}^n \right) - \frac{1}{2\mu(1-v)} \frac{k_1 k_2}{|\mathbf{k}|^4} \left[ k_1^2 \tilde{M}_{11}^n + k_2^2 \tilde{M}_{22}^n \right]. \tag{B30}$$

These conclude the update of $\tilde{\varepsilon}_{ij}(\mathbf{k})$ and hence $\varepsilon_{ij}(\mathbf{x})$.

We next calculate the detailed terms in the first two PDEs, where we are left with

$$\frac{\partial f_2}{\partial \phi_1} = -C_{ijkl} \left( \varepsilon_{ij} - \varepsilon_{ij}^*(\phi_1, \phi_2) \right) \frac{\partial \varepsilon_{kl}^*}{\partial \phi_1} = -C_{ijkl} \left( \varepsilon_{ij} - \varepsilon_{ij}^*(\phi_1, \phi_2) \right) \left( -\phi_2 \varepsilon_{kl}^{*1} + \varepsilon_{kl}^{*2} \right), \tag{B31}$$

$$\frac{\partial f_2}{\partial \phi_2} = -C_{ijkl} \left( \varepsilon_{ij} - \varepsilon_{ij}^*(\phi_1, \phi_2) \right) \frac{\partial \varepsilon_{kl}^*}{\partial \phi_2} = -C_{ijkl} \left( \varepsilon_{ij} - \varepsilon_{ij}^*(\phi_1, \phi_2) \right) (1-\phi_1) \varepsilon_{kl}^{*1}. \tag{B32}$$

We identify the stress field in the system $\sigma_{kl} = C_{ijkl} \left( \varepsilon_{ij} - \varepsilon_{ij}^*(\phi_1, \phi_2) \right)$, such that

$$\frac{\partial f_2}{\partial \phi_1} = \sigma_{kl} \left( \phi_2 \varepsilon_{kl}^{*1} - \varepsilon_{kl}^{*2} \right) = \sigma_{11} \left( \phi_2 \varepsilon^{*1} - \varepsilon^{*2} \right) + \sigma_{22} \left( \phi_2 \varepsilon^{*1} + \varepsilon^{*2} \right), \tag{B33}$$

$$\frac{\partial f_2}{\partial \phi_2} = \sigma_{kl} (\phi_1 - 1) \varepsilon_{kl}^{*1} = (\phi_1 - 1) \varepsilon^{*1} (\sigma_{11} + \sigma_{22}), \tag{B34}$$

where

$$\begin{cases} \sigma_{11} = \dfrac{2\mu}{1-2v} \left[ (1-v)\varepsilon_{11} + v\varepsilon_{22} \right] - M_{11}, \\ \sigma_{22} = \dfrac{2\mu}{1-2v} \left[ v\varepsilon_{11} + (1-v)\varepsilon_{22} \right] - M_{22}. \end{cases} \tag{B35}$$

Note that $\sigma_{12}$ is generally nonzero but did not appear in our expression. These conclude all the detailed expressions involved in the model algorithm.

**Movie 1.** The kinetics of photochemical phase transition at $T/T_m = 0.4$ with a fixed light intensity $\Gamma = 1.6$.

**Movie 2.** The kinetics of photochemical phase transition at $T/T_m = 0.8$ with a fixed light intensity $\Gamma = 1.6$.